\documentclass[9.5pt,journal,final,finalsubmission,twoside]{IEEEtran}

\usepackage[english]{babel}
\usepackage{amsmath,amssymb}
\usepackage{times}
\usepackage{relsize}
\usepackage{graphicx}
\usepackage{url}
\usepackage{booktabs}
\usepackage{multirow}
\usepackage{mparhack}
\usepackage{subfigure}
\usepackage{authblk}
\usepackage{amsthm}

 \usepackage{algorithm}
\usepackage{algorithmicx}
 \usepackage{algpseudocode}

\usepackage{epstopdf}

\usepackage{array}
\usepackage{cite}
\usepackage{eqparbox}
\usepackage{mdwmath}
\usepackage{balance}
\usepackage{epsfig}
\usepackage{xcolor}

\usepackage{mathtools}
\usepackage{bm}
\usepackage{amsfonts}
\usepackage[all]{xy}
\usepackage{etoolbox}
\usepackage{graphicx}
\DeclareMathAlphabet{\mathantt}{OT1}{antt}{li}{it}
\DeclareMathAlphabet{\mathpzc}{OT1}{pzc}{m}{it}

\usepackage{accents}
\DeclareFontFamily{OT1}{pzc}{}
\DeclareFontShape{OT1}{pzc}{m}{it}%
  {<-> s * [1.1] pzcmi7t}{}
\DeclareMathAlphabet{\mathpzc}{OT1}{pzc}%
                     {m}{it}

\DeclareMathOperator{\argmin}{\arg\min}

\makeatletter
\patchcmd{\@maketitle}
  {\addvspace{0.5\baselineskip}\egroup}
  {\addvspace{-1.45\baselineskip}\egroup}
  {}
  {}
\makeatother

\title{Improved Sparse Vector Code Based on Optimized Spreading Matrix for Short-Packet URLLC in mMTC}

\author{Linjie~Yang,~\IEEEmembership{Student Member,~IEEE},Pingzhi~Fan,~\IEEEmembership{Fellow,~IEEE}
	\thanks{Linjie Yang and Pingzhi Fan are with the School of Information Science and Technology, Southwest Jiaotong University, Chengdu 610031, China (e-mail: yanglinjie@my.swjtu.edu.cn; pzfan@swjtu.edu.cn).}
}

\begin{document}

\maketitle

\begin{abstract}
Recently, the sparse vector code (SVC) is emerging as a promising solution for  short-packet transmission in massive machine type communication (mMTC) as well as ultra-reliable and low-latency communication (URLLC). 
In the SVC process, the encoding and decoding stages are jointly
modeled as a standard compressed sensing (CS) problem. 
Hence, 
this paper aims at
improving the decoding performance of SVC by optimizing the spreading matrix (i.e. measurement matrix in CS).
To this end, two greedy algorithms to minimize the mutual coherence value of the spreading matrix in SVC are proposed.
Specially, for practical applications, the spreading matrices are further required to be bipolar whose entries are constrained as +1 or -1. As a result, the optimized spreading matrices are highly efficient for storage, computation, and hardware realization.
Simulation results reveal that,
compared with the existing work, 
the block error rate (BLER) performance of SVC can be improved significantly with the optimized spreading matrices.

\end{abstract}


\IEEEpeerreviewmaketitle

\section{Introduction}
To support seamless connection and reliable real-time interaction, 
short-packet transmission (i.e. sensor information and control information) plays a more and more important role in massive machine type communication (mMTC) as well as ultra-reliable and low-latency communication (URLLC) scenarios \cite{series2015imt}. 
Generally, reliable data transmission is realized through channel coding techniques. However, in most prior studies, the researchers are mostly focused on the long block coding to approach the Shannon theoretical bound.
To deal with this problem, the sparse vector code (SVC) with good block error rate (BLER) performance is proposed \cite{ji2018sparse} where the encoding and decoding process are modeled as a standard compressed sensing (CS) problem.
To improve the performance of SVC, an enhanced SVC (ESVC) is proposed in \cite{kim2020enhanced} where the modulated  M-quadrature amplitude modulation (QAM) data symbol is introduced to provide a higher freedom degree. 
Since non-zero elements are generated from the same QAM constellation alphabet, the received signal in ESVC may be null.
To remedy this, in \cite{zhang2021sparse}, the non-zero elements are generated via QAM with constellation rotation (CR).

Different from \cite{kim2020enhanced,zhang2021sparse}, 
this paper aims at improving the decoding performance of SVC by optimizing the spreading matrix, or the measurement matrix in CS as the CS plays a core role in SVC.
In fact, measurement matrix optimization is widely studied in the CS field.
In \cite{elad2007optimized}, measurement matrix optimization is firstly proposed by Elad, where the average mutual coherence of the measurement matrix is optimized by a shrinkage-based method.
In \cite{tropp2005designing}, an equiangular tight frame (ETF) based method is proposed whose objective is to find an equivalent matrix which is mostly close to an ETF.
Based on \cite{tropp2005designing}, a gradient-based alternating minimizing approach is developed to obtain the optimized measurement matrix in \cite{abolghasemi2012gradient}. In \cite{yan2014shrinkage}, the authors combine the shrinkage-based method in \cite{elad2007optimized} and alternating projection technique in \cite{tropp2005designing} to jointly optimize the measurement matrix, which leads to a better signal reconstruction performance.
However, these measurement matrix optimization methods are implemented in a complex number field, which results in inefficiency in storage and hardware realization in wireless communication scenarios. As a result, the bipolar measurement matrix, whose entries are required to be +1 or -1, attracts much attention.

In \cite{ji2018sparse}, a bipolar measurement matrix with dimension $\mathbf{C}\in\{+1,-1\}^{L_{s}\times L_{sv}}$ is obtained by simply mapping the zero elements of a random Bernoulli matrix into $-1$. 
In \cite{li2012convolutional}, a cyclic matrix with dimension $\mathbf{H}\in\{+1,-1\}^{N\times N}, L_{sv}\leq N$ is firstly obtained by cyclically shifting an m-sequence with length $N$. Then, $L_{s}$ rows and $L_{sv}$ columns of the cyclic matrix are randomly taken as the measurement matrix. Hence, the constructed measurement matrix is referred to as a partial M (PM) matrix. For simplicity, we assume $N=L_{sv}$. Meanwhile, we denote the other $N-L_{s}$ rows which are not selected into the measurement matrix as the redundant rows of the cyclic matrix $\mathbf{H}$.
Equivalently, the PM measurement matrix could be formed by deleting the redundant rows of the cyclic matrix $\mathbf{H}$ randomly.
In \cite{gan2019bipolar}, a chaotic sequence based bipolar measurement matrix (CBM) is constructed.
Similar to the PM method, a partial Hadamard matrix (PHM) based measurement matrix is constructed in \cite{bryant2017compressed}.
In \cite{yu2020binary}, a binary Golay sequence based bipolar measurement matrix (BGM) with excellent CS performance is developed.

Inspired by the above studies, the optimization of bipolar measurement matrix is also considered based on PHM. Our main contributions can be summarized as follows.
\begin{itemize}
	\item 
	Firstly, an optimized partial Hadamard matrix (OPHM) is proposed. 
    Different from \cite{bryant2017compressed} where the author randomly deletes the redundant rows of a Hadamard matrix $\mathbf{H}\in\{+1,-1\}^{L_{sv}\times L_{sv}}$, in our method, the redundant rows are deleted iteratively in a locally optimal order. 
     
	\item 
	Secondly, an optimized column augmentation (OCA) algorithm is proposed, where the spreading matrix is generated column by column. In each iteration, the mutual coherence value between the newest generated column and  all  previously generated columns stored in spreading matrix, is minimized. 
%	Such an iteration is further modeled as a standard binary integer programming (BIP) problem and is solved efficiently by off-the-shelf integer programming (IP) optimization methods, owing to the small dimension of the spreading matrix in SVC \cite{ji2018sparse,zhang2021sparse,kim2020enhanced}.
	
	\item 
	 It is shown by simulation results that, compared with the existing work in 
	 \cite{ji2018sparse,li2012convolutional,gan2019bipolar,yu2020binary,bryant2017compressed}, the block error rate (BLER) of SVC can be improved significantly with our optimized spreading matrices for Short-Packet URLLC in mMTC.
	 
\end{itemize}

The rest of the paper is organized as follows. The system model is introduced in Section II. The proposed bipolar spreading matrix optimization methods are given in Section III. 
Simulation results are demonstrated in Section IV. Finally, the paper is
concluded in Section V

\section{System Model}

\begin{figure}[htbp]
	\centering
	\includegraphics[scale=0.95]{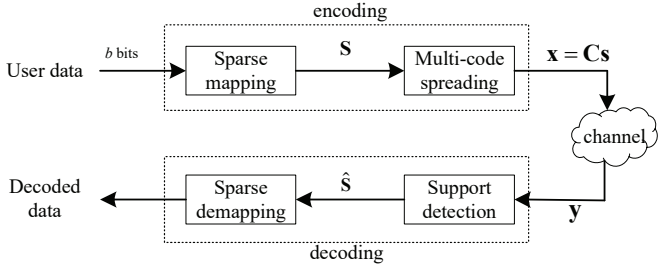}
	\caption{The block diagram of SVC.}
	\label{1}
\end{figure}

As shown in Fig. \ref{1}, in SVC, $b$ bits user data are firstly mapped into the non-zero elements' indices of a $K-$sparse vector $\mathbf{s}$ with length $L_{sv}$. 
The operation is referred to as sparse mapping.
To enhance the sparsity,
the value of $K$ is generally small (i.e. $K=2$ in \cite{ji2018sparse,zhang2021sparse}). Hence, the minimal length of the sparse vector $\mathbf{s}$, $L_{sv}$ can be computed through the following equation. 
\begin{equation}
\lfloor \log_{2}(\binom{K}{L_{sv}}) \rfloor \geq b.
\end{equation}
where $\lfloor \cdot \rfloor$ denotes the round down operation.
For example, the process of sparse mapping of $5$ bits user data is illustrated in formula (2) where $L_{sv} = 9, K =2$. More details of the sparse mapping rules can be found in the table I of \cite{ji2018sparse}.
\begin{equation}
\begin{aligned}
&00000 \quad \longleftrightarrow \quad 000000011\\
&00001 \quad \longleftrightarrow \quad  000000101\\
&00011 \quad \longleftrightarrow \quad 000001001 \\
&\quad \vdots \quad\quad\quad\ \vdots \quad\quad\quad\quad \vdots \\
&11111 \quad \longleftrightarrow \quad 100000001.
\end{aligned}
\end{equation}
After the sparse mapping, the transmitted signal $\mathbf{x}$ is generated by the multi-code spreading operation. Concretely, let $\mathbf{C} = [\mathbf{c}_{1},\mathbf{c}_{2},\cdots,\mathbf{c}_{L_{sv}}]$ denotes a spreading matrix, where $\mathbf{c}_{i} = [ \mathbf{c}[1,i],\mathbf{c}[2,i],\cdots,\mathbf{c}[L_{s},i]  ]^{T}$ is the $i^{th}$ spreading sequence with length $L_{s}$. $(\cdot)^{T}$ denotes the transpose operation of a vector. The generation of the transmitted signal $\mathbf{x}$ in multi-code spreading process can be written as 
\begin{equation}
\mathbf{x} = \mathbf{C}\mathbf{s},
\end{equation}
In \cite{ji2018sparse,zhang2021sparse}, the elements of the spreading matrix $\mathbf{C}$ are sampled from a Bernoulli distribution. Example of $\mathbf{C}$ for $L_{s}=5$ and $L_{sv}=9$ is given by
\begin{equation}
\footnotesize
\mathbf{C}=\frac{1}{\alpha}\left[\begin{array}{rrrrrrrrr}
1&1&1&1&-1&1&-1&1&-1 \\
1&-1 & 1 & -1 & 1 & -1 & 1 & -1 & -1\\
1&1 & -1 & -1 & 1 & 1 & -1 & -1 & 1\\
1&-1 & -1 & 1 & 1 & -1 & 1 & 1 & -1\\
-1&1 & 1 & 1 & -1 & -1 & -1 & -1 & 1 
\end{array}\right].
\end{equation}
where $\alpha$ is the normalization factor to normalize the unit power of the transmitted signal $\mathbf{x}$.

At the base station (BS), considering the channel fading, the received signal $\mathbf{y}$ is given by
\begin{equation}
\mathbf{y} = \mathbf{h} \circ \mathbf{x}+\mathbf{n},
\end{equation}
where $\mathbf{h}=[\mathbf{h}[1],\mathbf{h}[2],\cdots,\mathbf{h}[L_{s}]]^{T}$ denotes the channel coefficient vector between user and BS.
$\circ$ denotes the element-wise product of two vectors.
$\mathbf{h}[l]=1,1\leq l \leq L_{s}$ under Gaussian channel and $\mathbf{h}[l] \sim \mathcal{CN}(0,\sigma_{R}^{2}), 1\leq l \leq L_{s}$ under Rayleigh channel. $\mathbf{n}$ denotes the Gaussian background noise which obeys $\mathcal{CN}(0,\sigma_{G}^{2})$.
Apparently, recover $\mathbf{s}$ from $\mathbf{y}$ is a standard CS problem. 
According to \cite{ji2018sparse,zhang2021sparse,kim2020enhanced}, multi-path match pursuit (MMP) algorithm is employed for support detection $\hat{\mathbf{s}}$ in this paper\footnote{Support denotes the set of the indices of non-zero elements in $\mathbf{s}$. For example, $\mathbf{s}=[1,0,0,0,1]$, the support $\Omega_{\mathrm{s}}=\{1,5\}$.}.
Finally, the user data information is decoded from $\hat{\mathbf{s}}$ by spare demapping operation which can be implemented by looking up the relationship table given in (2).
Additionally, as reported in \cite{ji2018sparse,zhang2021sparse}, quadrature phase shift keying (QPSK) modulation is also adopted in this paper.

\section{Bipolar spreading Matrix Optimization}
As stated in \cite{ji2018sparse,kim2020enhanced,zhang2021sparse}, the decoding performance of SVC is significantly affected by the mutual coherence of the spreading matrix $\mathbf{C}$. 
The mutual coherence \cite{elad2007optimized,tropp2005designing,abolghasemi2012gradient,yan2014shrinkage} of $\mathbf{C}$ is defined as
\begin{equation}
\mu (\mathbf{C}) = \max_{1 \leq i \neq j \leq L_{sv}} \frac{ \vert <\mathbf{C}[:,i],\mathbf{C}[:,j]> \vert }{ || \mathbf{C}[:,i] ||_{2}\cdot ||\mathbf{C}[:,j]||_{2}   }
\end{equation}
where $\vert\cdot\vert$ returns the absolute value. $||\cdot||_{2}$ denotes the $l_{2}$-norm. $\mathbf{C}[:,i]$ denotes the $i^{th}$ column of $\mathbf{C}$. 
$\mathbf{C}[i,:]$ denotes the $i^{th}$ row of $\mathbf{C}$. 
$\mu(\mathbf{C})$ is also the largest off-diagonal element of the Gram matrix of $\mathbf{C}$, $\mathbf{G}$ (i.e. $\mathbf{G}[i,j], i\neq j$), which is computed as $\mathbf{G}=\tilde{\mathbf{C}}^{T}\tilde{\mathbf{C}}$.
$\tilde{\mathbf{C}}$ is obtained by normalizing each column of $\mathbf{C}$.
Since the elements of the spreading matrix $\mathbf{C}$ are +1 or -1, the $l_{2}$ norm values of any two columns are the same in (6), i.e.   $||\mathbf{C}[:,i]||_{2}=||\mathbf{C}[:,j]||_{2}= \sqrt{L_{s}},1\leq i\neq j \leq L_{sv}$. Hence, the spreading matrix optimization in (6) can be simplified as
\begin{equation}
\begin{array}{ll}
\mathop{\min} \limits_{\mathbf{C}} & \mathop{\max} \limits_{ 1\leq i \neq j \leq L_{sv} } \vert  < \mathbf{C}[:,i], \mathbf{C}[:,j] >\vert\\
\text { s.t. } & \mathbf{C}\in \{-1,+1\}^{L_{s}\times L_{sv}} \\
\end{array}
\end{equation}
In \cite{elad2007optimized}, Elad declares the problem in (7) is NP-hard. 
Hence, similar to the existing studies\cite{li2012convolutional,gan2019bipolar,bryant2017compressed,yu2020binary,elad2007optimized,yan2014shrinkage,tropp2005designing,abolghasemi2012gradient}, 
the objective of this paper is to find a sub-optimal solution.

\subsection{Optimized Partial Hadamard Matrix}
In \cite{bryant2017compressed}, PHM is proposed where $L_{s}$ rows and $L_{sv}$ columns are randomly taken from a Hadamard matrix with a larger dimension.  It is due to the orthogonal and non-linear related characteristics of the Hadamard matrix, the CS reconstruction performance of PHM is excellent. However, 
the author ignores that different rows may have different contributions to the mutual coherence of the Hadamard Matrix. 
Hence, the PHM method can be further improved.

Concretely, in our method, the redundant rows of the Hadamard matrix are removed iteratively. In each iteration, the row which leads the minimal mutual coherence value of the remaining Hadamard matrix is deleted. 
If there are multiple such rows, we would delete any one of them randomly.
Let $\mathbf{C}^{(i)}$ denote the remaining Hadamard matrix in the $i^{th}$ iteration. $\mathbf{C}[ \backslash \{i\}, : ]$ denotes removing the $i^{th}$ row of the spreading matrix $\mathbf{C}$. 
Hence, the $i^{th}$ iteration can be summarized as
\begin{equation}
\begin{aligned}
&\hat{k} = \mathop{\argmin}_{ 1\leq k \leq L_{sv}-i } \mu( \mathbf{C}^{(i-1)}[ \backslash\{k\},: ] ),\\
&\mathbf{C}^{(i)} =  \mathbf{C}^{(i-1)}[ \backslash\{\hat{k}\},: ]. 
\end{aligned}
\end{equation}
Initially, $\mathbf{C}^{(0)} \in \{+1,-1\}^{L_{sv}\times L_{sv}}$ is constructed by the first $L_{sv}$ rows and first $L_{sv}$ columns of a Hadamard matrix $\mathbf{H} \in \{+1,-1\}^{N\times N}, N\geq L_{sv}$. 
The pseudo-code of the proposed OPHM is given in Algorithm 1.
\begin{algorithm}[htbp]
	\caption{The proposed OPHM algorithm}
	\begin{algorithmic}[1] 
		\Require $L_{s}, L_{sv}$
	    \Ensure $\mathbf{C} \in \{+1,-1\}^{L_{s}\times L_{sv}}$	
		\State Construct a Hadamard matrix $\mathbf{H}\in \{+1,-1\}^{N \times N}, N\geq L_{sv}$.
		\State Initialize $\mathbf{C}^{(0)}$ by the first $L_{sv}$ rows and columns of $\mathbf{H}$.
		\For {$i=1:L_{sv}-L_{s}$}
		\State Obtain $\mathbf{C}^{(i)}$ according to formula (8).
		\EndFor
		\State $\mathbf{C} = \mathbf{C}^{(L_{sv}-L_{s})}$.
	\end{algorithmic}
\end{algorithm}

\subsection{Optimized Column Augmentation}
Moreover, the optimized column augmentation (OCA) algorithm is proposed. Instead of solving the problem in (7) directly, in the OCA algorithm,
the spreading matrix is optimized column-by-column.

To be more concrete, 
in the $i^{th}$ iteration, a new column $\mathbf{c}\in \{+1,-1\}^{L_{s}\times 1}$ is generated. The maximum of the inner product value between $\mathbf{c}$ and the previously generated columns $\mathbf{C}[:,j], 1\leq j \leq i-1$ is minimized. Then, $\mathbf{c}$ is added into $\mathbf{C}$ to be its $i^{th}$ column. The process of the $i^{th}$ iteration can be modeled as
\begin{equation}
\begin{array}{ll}
\mathop{\min}\limits_{\mathbf{c}} & \mathop{\max} \limits_{ 1\leq j \leq i-1 } \vert  < \mathbf{c}, \mathbf{C}[:,j] >\vert\\
\text { s.t. } & \mathbf{c}\in \{-1,+1\}^{L_{s}\times 1} \\
\end{array}
\end{equation}
Here, let $m^{(i)} = \mathop{\max} \limits_{ 1\leq j \leq i-1 } \vert  < \mathbf{c}, \mathbf{C}[:,j] >\vert$.
It is notable  that the value of $m^{(i)}$ is a monotone non-decreasing positive integer, i.e. $m^{(i)}\geq m^{(i-1)}\in \mathbb{Z}_{+}$. The minimal increment of $m^{(i)}$ could be 1 in our case. Hence, the value of $m^{(i)}$ is estimated independently by utilizing an incremental approach.

Then, the problem in (9) can be largely simplified once $m^{(i)}$ is given.
\begin{equation}
\begin{array}{ll}
\mathop{\min}\limits_{\mathbf{c}} & m^{(i)} \\
\text { s.t. } & -m^{(i)} \leq\mathbf{c}^{T}\cdot\mathbf{C}[:,j]\leq m^{(i)} \\
& \mathbf{c}\in \{-1,+1\}^{L_{s}\times 1} \\
&  1\leq j \leq i-1
\end{array}
\end{equation}
The objective now becomes finding a feasible solution of the problem in (10). This problem can be solved by the interior-point method \cite{mehrotra1992implementation}. To this end, we convert the problem in (10) equivalently into (11)
\begin{equation}
\begin{array}{ll}
\mathop{\min}\limits_{\mathbf{c}} & \sum_{j=1}^{i-1} u( <\mathbf{C}[:,j],\mathbf{c}> - m^{(i)} )+\\
 &  \sum_{j=1}^{i-1} u( -<\mathbf{C}[:,j],\mathbf{c}> - m^{(i)} )\\
\text { s.t. } 
& \mathbf{c}\in \{-1,+1\}^{L_{s}\times 1} \\
\end{array}
\end{equation}
where $u(\cdot)$ denotes an indicator function where $u(x)=0,x\leq 0$, otherwise $u(x) = \infty$.
Since the function $u(x)$ is not differentiable,
in the practical solving process, the function $u(x)$ is approximated by the function $u_{t}(x)$.
\begin{equation}
u_{t}(x)= \begin{cases} -\frac{1}{t}\log(-x), &  x < 0,\\ 
0, &x =0,\\
\infty, &  { \rm otherwise }\end{cases}
\end{equation}
where the constant $t$ determine the degree of approximation of $u(x)$ by $u_{t}(x)$.
Finally, the problem in (11) is further converted into (13) by variable substitution $\mathbf{c}=2\mathbf{g}-1$.
\begin{equation}
\begin{array}{ll}
\mathop{\min}\limits_{\mathbf{g}} & f^{(i)}\\
\text { s.t. } 
& \mathbf{g}\in \{0,1\}^{L_{s}\times 1} \\
\end{array}
\end{equation}
where $  f^{(i)} = -\frac{1}{t}\sum_{j=1}^{i-1} \log( -<\mathbf{C}[:,j],2\mathbf{g}-1> + m^{(i)} )-\frac{1}{t}\sum_{j=1}^{i-1} \log( <\mathbf{C}[:,j],2\mathbf{g}-1> + m^{(i)} )   $.
Apparently, $f^{(i)}$ is a convex function, hence, the problem in (13) is a standard binary integer programming (BIP) problem and can be efficiently solved by off-the-shelf integer programming methods, e.g. branch-and-bound or branch-and-price \cite{liuzzi2022computational}.
Additionally, $t=100$, $\mathbf{C}[:,1]=[1,1,\cdots,1]^{L_{s}\times 1}$ is initialized.
The pseudo-code of the proposed OCA algorithm is given in Algorithm 2.

It is noted that in the existing works \cite{ji2018sparse,li2012convolutional,gan2019bipolar,yu2020binary,bryant2017compressed}, the measurement matrices are generally constructed based on certain algebraic structures. 
Hence, the computational complexities of these methods are lower than that of the proposed OPHM and OCA algorithms.
However, the expense of computational complexity in spreading matrix $\mathbf{C}$ construction is not a concern in SVC for two reasons.
Firstly, the authors of \cite{zhang2021sparse,ji2018sparse} reveal that the SVC  is highly efficient only when the number of information bits is small, i.e. $b \leq 12$.
It implies that the dimension of the spreading matrix $\mathbf{C}$ in SVC is normally very small, e.g. $b=12, L_{sv} = 92$, $r= \frac{L_{s}}{L_{sv}}=\frac{1}{2}$, thus $L_{s} = 46$, $\mathbf{C}\in\{+1,-1\}^{46\times 92}$.
Secondly, in SVC, the spreading matrix $\mathbf{C}$ 
can be optimized in an offline manner and then stored at the BS in advance.
As a result, the complexity of the practical applications of the SVC would not be increased.

\begin{algorithm}[htbp]
	\caption{The proposed OCA algorithm}
	\begin{algorithmic}[1] 
		\Require $L_{s}, L_{sv}$
		\Ensure $\mathbf{C} \in \{+1,-1\}^{L_{s}\times L_{sv}}$. 	
		\State Initialize $\mathbf{C}[:,1]=[1,1,\cdots,1]^{L_{s}\times 1}$ ,  $i=1$.
		\State $m^{(i)}=0$.
		\While{$i<L_{sv}$}
		\State Obtain $\mathbf{g}$ by solving the problem in (13).
		\If{$f^{(i)}=\infty$}
		\State $m^{(i)}=m^{(i)}+1$.
		\Else
		\State $i=i+1$.
		\State $\mathbf{C}[:,i]=2\mathbf{g}-1$.
		\State $m^{(i)}=m^{(i-1)}$.
		\EndIf
		\EndWhile
	\end{algorithmic}
\end{algorithm}

\section{Simulation Results}
In this section, the block error rate (BLER) performances of the
SVC algorithm utilizing different spreading matrix $\mathbf{C}$ \cite{ji2018sparse,li2012convolutional,gan2019bipolar,yu2020binary,bryant2017compressed} are simulated. The parameters of system configuration are listed in the table I.
\begin{table}[htbp]
	\caption{System Configuration}
	\begin{center}
		\begin{tabular}{ c|c} 
			\hline
      The value of $L_{sv}$  &24, 65\\
			\hline
		 the value of $L_{s}$     &   15, 25, 40 \\
			\hline
		The value of $K$  &2 \\
			\hline
		Data packet length $b$ & 8, 11 \\
			\hline
	   The variance of Rayleigh fading $\sigma_{R}^{2}$ & 1\\
			\hline
		\end{tabular}
	\end{center}
\end{table}

\begin{figure}
	\centering %表示居中
	\includegraphics[scale=0.25]{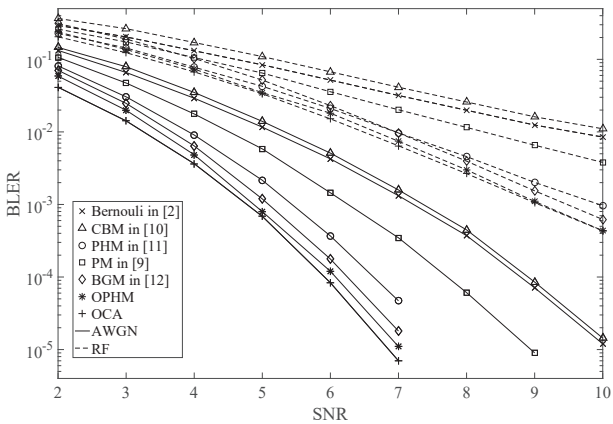}
	\caption{BLER performance comparison between the proposed spreading matrices and the counterparts in \cite{ji2018sparse,li2012convolutional,gan2019bipolar,yu2020binary,bryant2017compressed}, where $\mathbf{C}\in \{+1,-1\}^{15\times 24}$.}
	%图片的名称
	\label{2}
	%图片的标签，用于文章中的引用，注意到标签的数字与实际文章显示的数字可能不同
\end{figure}

The BLER performances of the optimized spreading matrices with the dimension $\mathbf{C}\in\{+1,-1\}^{15\times 24}$ is presented in Fig. \ref{2}. 
In this condition, the length of data packet is $8$ bits. The sampling ratio is $r =\frac{5}{8}$.
The BLER performances of the proposed OPHM and OCA outperform that of the BGM in \cite{yu2020binary} which has the best BLER performance in the existing counterparts.
The OCA algorithm achieves the best BLER performance.
When compared with the BGM algorithm, OCA achieves around 0.38 dB under Gaussian channel at the target accuracy of $10^{-5}$ and around 0.48 dB under Rayleigh channel at the target accuracy of $10^{-3}$.
When compared with the Bernoulli spreading matrix in the original SVC in \cite{ji2018sparse}, the performance gain exceeds 3 dB regardless of the channel type.

\begin{figure}
	\centering %表示居中
	\includegraphics[scale=0.25]{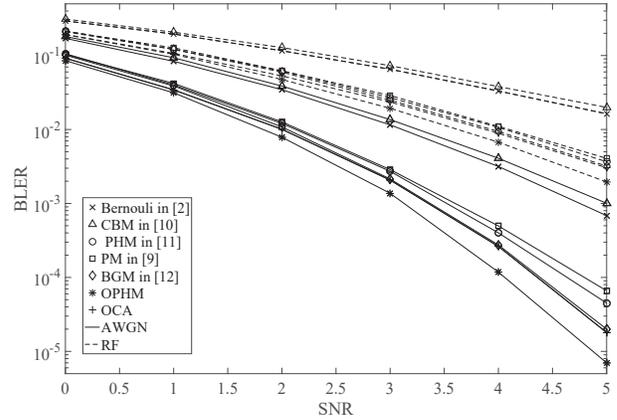}
	\caption{BLER performance comparison between the proposed spreading matrices and the counterparts in \cite{ji2018sparse,li2012convolutional,gan2019bipolar,yu2020binary,bryant2017compressed}, where $\mathbf{C}\in \{+1,-1\}^{20\times 24}$.}
	%图片的名称
	\label{3}
	%图片的标签，用于文章中的引用，注意到标签的数字与实际文章显示的数字可能不同
\end{figure}

The BLER performance of the optimized spreading matrices with the dimension 
$\mathbf{C}\in \{+1,-1\}^{20\times24}$
 is presented in Fig. \ref{3}. 
The sampling ratio is $r =\frac{5}{6}$.
In this condition, the BLER performance of the BGM algorithm \cite{yu2020binary} is almost the same as that of the proposed OCA algorithm.
The OPHM algorithm achieves the best BLER performance.
When compared with the BGM algorithm, the OPHM achieves around 0.38 dB under Gaussian channel at the target accuracy of $10^{-5}$ and around 0.5 dB under Rayleigh channel at the target accuracy of $10^{-2}$. 
When compared with the Bernoulli spreading matrix in \cite{ji2018sparse}, the performance gain of OPHM is around 2 dB regardless of the channel type.

\begin{figure}
	\centering %表示居中
	\includegraphics[scale=0.25]{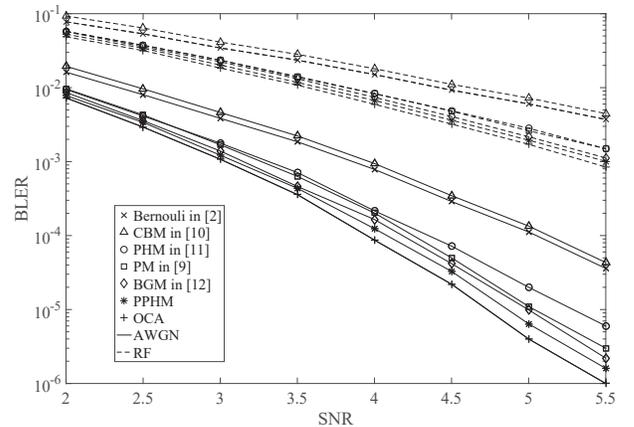}
	\caption{BLER performance comparison between the proposed spreading matrices and the counterparts in \cite{ji2018sparse,li2012convolutional,gan2019bipolar,yu2020binary,bryant2017compressed}, where $\mathbf{C}\in \{+1,-1\}^{25\times 65}$.}
	%图片的名称
	\label{4}
	%图片的标签，用于文章中的引用，注意到标签的数字与实际文章显示的数字可能不同
\end{figure}

The BLER performance of the optimized spreading matrices with larger dimension $\mathbf{C}\in \{+1,-1\}^{25\times 65}$ is also tested.
The length of data packet $b$ becomes $11$ bits. The sampling ratio is $r=\frac{5}{13}$.
At this time, the proposed OCA algorithm achieves the best BLER performance.
When compared with the BGM scheme \cite{yu2020binary}, the performance gain of OCA is around 0.36 dB under Gaussian channel at the target accuracy of $10^{-6}$ and 0.26 dB under Rayleigh channel at the target accuracy of $10^{-3}$.
When compared with the Bernoulli spreading matrix in \cite{ji2018sparse}, 
the performance gain of OCA is around 1.5 dB regardless of the channel type.

\begin{figure}
	\centering %表示居中
	\includegraphics[scale=0.25]{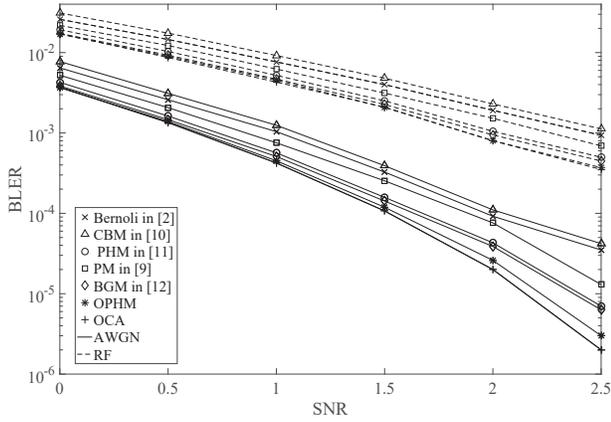}
	\caption{BLER performance comparison between the proposed spreading matrices and the counterparts in \cite{ji2018sparse,li2012convolutional,gan2019bipolar,yu2020binary,bryant2017compressed}, where $\mathbf{C}\in \{+1,-1\}^{40\times 65}$.}
	%图片的名称
	\label{5}
	%图片的标签，用于文章中的引用，注意到标签的数字与实际文章显示的数字可能不同
\end{figure}

The BLER performance of the optimized spreading matrices with the dimension $\mathbf{C}\in \{+1,-1\}^{40\times 65}$ is presented in Fig. \ref{5}.
Again, the proposed OCA algorithm achieves the best BLER performance.
When compared with the BGM algorithm \cite{yu2020binary}, the performance gain of OCA is around 0.25 dB under Gaussian channel at the target accuracy of $10^{-6}$ and 0.2 dB under Rayleigh channel at the target accuracy of $10^{-2}$.
When compared with the Bernoulli spreading matrix \cite{ji2018sparse}, 
the performance gain of OCA is around 1 dB regardless of the channel type.

\begin{figure}
	\centering %表示居中
	\includegraphics[scale=0.29]{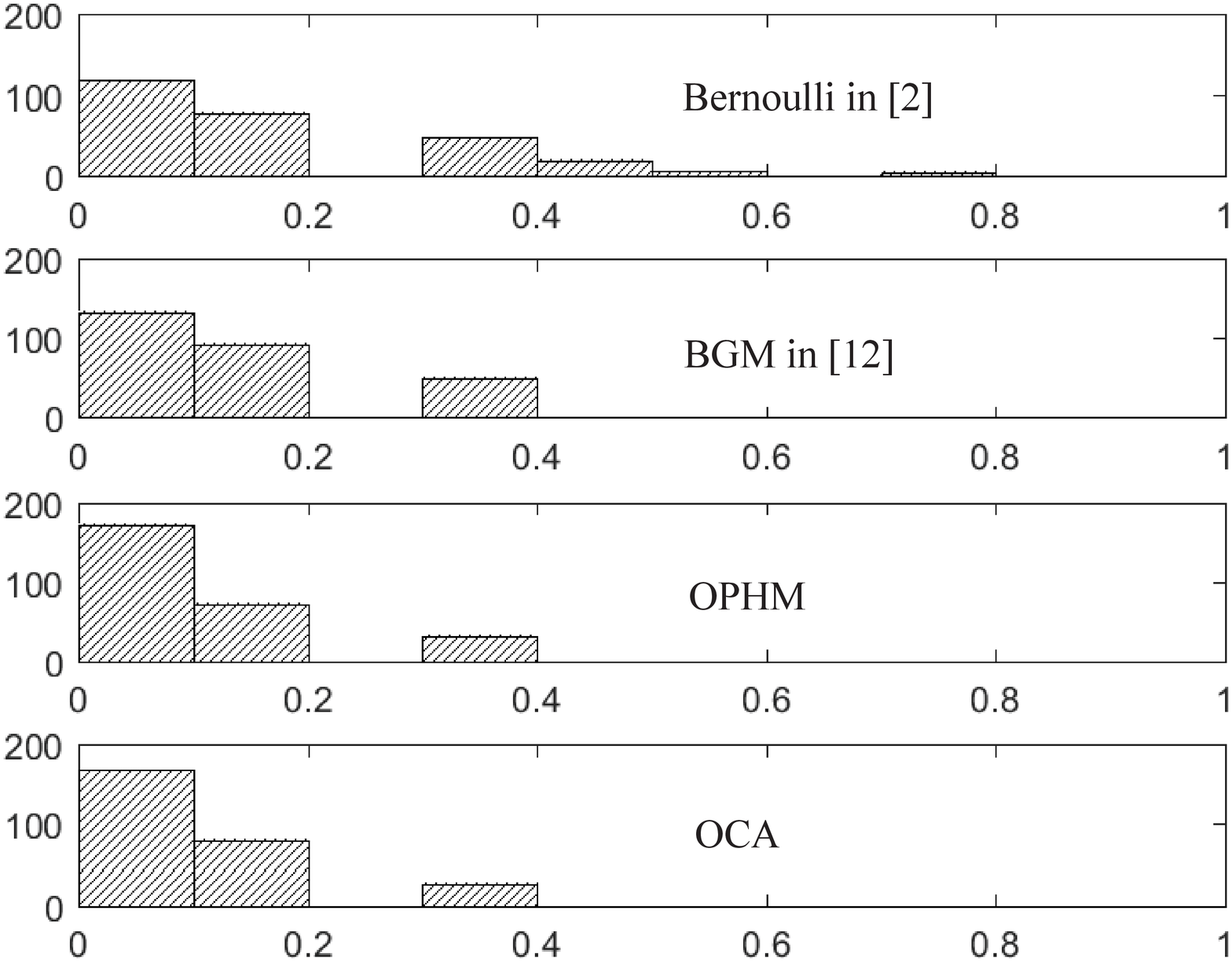}
	\caption{Histogram of the absolute off-diagonal entries of Gram matrix $\mathbf{G}$ of the optimized spreading matrices and the counterparts in \cite{ji2018sparse,yu2020binary}, where $\mathbf{C}\in \{+1,-1\}^{15\times 24}$.}
	%图片的名称
	\label{6}
	%图片的标签，用于文章中的引用，注意到标签的数字与实际文章显示的数字可能不同
\end{figure}
Finally, the histogram of the absolute off-diagonal entries of the gram matrix $\mathbf{G}$, which can reflect the mutual coherence distribution of the spreading matrix $\mathbf{C}$ vividly, is illustrated in Fig. \ref{6}. 
Both the Bernoulli spreading matrix in the original SVC algorithm \cite{ji2018sparse} and the BGM algorithm\cite{yu2020binary} which has the best BLER performance in the existing counterparts are mainly concentrated.
Compared with the Bernoulli spreading matrix, the mutual coherences of BGM, OPHM and OCA are significantly reduced.
Compared with BGM \cite{yu2020binary}, the average mutual coherence \cite{elad2007optimized} of OPHM and OCA outperform BGM significantly.
Because, in the OPHM and OCA algorithms, the ratio of entries in $\mathbf{G}$ whose values lies in the range from 0.3 to 0.4 and the range from 0.1 to 0.2 are significantly lower than that of the BGM algorithm \cite{yu2020binary}.

\section{Conclusions}
In this paper, two greedy spreading matrix optimization algorithms are developed to minimize the mutual coherence of the spreading matrix in the SVC. In the OPHM algorithm, the spreading matrix is firstly initialized as a Hadamard matrix whose dimension is larger than or equal to $L_{sv}$. Then, the redundant rows of the spreading matrix are removed in an locally optimal order.
In the OCA algorithm, the spreading matrix is optimized column-by-column. 
Each iteration is formulated as a standard BIP problem which can be solved by  off-the-shelf integer programming methods. 
Simulation results verify the efficiency and superior performance of our proposal.

\bibliographystyle{IEEEtran}
\bibliography{IEEEabrv,ForIEEEBib}

\end{document}